\documentclass[aps,preprint]{revtex4}

\begin{document}

\title{Restricted sandpile revisited}

\author{Ronald Dickman$^{\dagger}$}
\address{
Departamento de F\'{\i}sica, ICEx, Universidade Federal de Minas
Gerais, Caixa Postal 702, 30161-970 Belo Horizonte, Minas Gerais,
Brazil}
\date{\today}

\begin{abstract}
I report large-scale Monte Carlo studies of a one-dimensional
height-restricted stochastic sandpile using the quasistationary
simulation method. Results for systems of up to 50$\,$000 sites
yield estimates for critical exponents that differ significantly
from those obtained using smaller systems, but are consistent with
recent predictions derived from a Langevin equation for stochastic
sandpiles [Ramasco et al., Phys. Rev. E{\bf 69}, 045105(R) (2004)].
This suggests that apparent violations of universality in
one-dimensional sandpiles are due to strong corrections to scaling
and finite-size effects.

\vspace{2em}

\noindent$^\dagger$email: dickman@fisica.ufmg.br
\end{abstract}




\date{\today}
\maketitle

\section{Introduction}

Sandpile models are the prime example of self-organized criticality
(SOC) \cite{btw,dhar99}, a control mechanism that forces a system
with an absorbing-state phase transition to its critical point
\cite{bjp,granada}, leading to scale invariance in the apparent
absence of parameters \cite{ggrin}. SOC in a slowly-driven sandpile
corresponds to an absorbing-state phase transition in a model having
the same local dynamics, but a fixed number of particles
\cite{bjp,tb88,pmb96,vz,dvz,vdmz}.  The latter class of models is
usually designated as {\it fixed-energy sandpiles} (FES) or {\it
conserved sandpiles}.  Continuous absorbing-state phase transitions
characterized by a nonconserved order parameter (activity density)
coupled to a conserved field that does not diffuse in the absence of
activity, are expected to define a universality class \cite{rossi}.
This class, referred to as C-DP (that is, a model-C version, in the
sense of Halperin and Hohenberg \cite{hh}, of directed percolation,
or DP), appears to be distinct from that of directed percolation
\cite{ramasco}.

In recent years considerable progress has been made in
characterizing the critical properties of conserved stochastic
sandpiles, although no complete, reliable theory is yet at hand.  As
is often the case in critical phenomena, theoretical understanding
of scaling and universality rests on the analysis of a continuum
field theory or Langevin equation (a nonlinear stochastic partial
differential equation) that reproduces the phase diagram and
captures the fundamental symmetries and conservation laws of the
system. Important steps in this direction are the recent numerical
studies of a Langevin equation \cite{ramasco,dornic} for C-DP.  (The
latter which appears to incorporate the essential aspects of
stochastic sandpiles.) The critical exponent values reported in Ref.
\cite{ramasco} are in good agreement with simulations of conserved
lattice gas (CLG) models \cite{pastor,kockelkoren}, which exhibit
the same symmetries and conservation laws as stochastic sandpiles.

The Langevin equation exponents are also consistent with the best
available estimates for stochastic sandpiles in two dimensions
\cite{ramasco}, with the exception of the exponent $\theta$
governing the initial decay of the order parameter.  (The
discrepancy regarding $\theta$ likely reflects strong corrections to
short-time scaling in sandpiles, due to long memory effects
associated with initial density fluctuations \cite{manna1d}.)
Pending a better understanding of this question, it appears that
stochastic sandpiles are consistent with C-DP in two dimensions. In
the one-dimensional case, however, there is a significant
discrepancy between the Langevin equation results and those for
sandpile models.

Specifically, analysis of the Langevin equation for C-DP yields, in
one dimension, the order-parameter critical exponent value $\beta =
0.28(2)$, while previous studies
\cite{manna1d,mnrst,mancam,lubeckcttp} of stochastic sandpiles
furnish values near 0.40 for this exponent. There are also smaller
discrepancies for other critical exponents. If this discrepancy were
to persist, one would be forced to conclude that the proposed
Langevin equation misses some essential aspect of sandpiles (at
least in the one-dimensional case), or that not all models with the
same symmetries and conserved quantities belong to the same
universality class. In an effort to clarify the situation, I apply
the recently devised quasistationary simulation method
\cite{qsspre,qss2,qss3} to the restricted-height sandpile introduced
in Ref. \cite{mnrst}.

The balance of this paper is organized as follows. In Sec. II I
define the model and summarize the simulation method. Numerical
results are analyzed in Sec. III, and in Sec. IV I discuss the
findings in the context of universality.

\section{Model}

I study the "independent" version of the model introduced in Ref.
\cite{mnrst}. The system, a continuous-time, restricted-height
version of Manna's stochastic sandpile \cite{manna}, is defined on a
ring of $L$ sites.  The configuration is specified by the number of
particles, $z_i = 0, 1$, or 2, at each site $i$. Sites with $z_i
\!=\! 2$ are {\it active}, and have a toppling rate of unity. The
continuous-time Markovian dynamics consists of a series of toppling
events at individual sites.  When site $i$ topples, two particles
attempt to move randomly (and independently) to either $i-1$ or
$i+1$. (The two particles may both try to jump to the same
neighbor.) Each particle transfer is accepted so long as it does not
lead to a site having more than two particles. (If the target site
is already doubly occupied the particle does not move. Thus an
attempt to send two particles from site $j$ to site $k$, with $z_k =
1$, results in $z_k\!=\!2$ and $z_j=1$.) The next site to topple is
chosen at random from a list of active sites, which is updated
following each event. The time increment associated with each
toppling is $\Delta t = 1/N_A$, where $N_A$ is the number of active
sites just prior to the event.

Any configuration devoid of doubly occupied sites is absorbing.
Although absorbing configurations exist for particle densities $p =
N/L \leq 1$, the critical value $p_c$ (above which activity
continues indefinitely) appears to be strictly less than unity. In
Ref. \cite{mnrst} the model was studied in the site and pair
mean-field approximation (which yield a continuous phase transition
at $p_c = 0.5$ and 0.75, respectively, in one dimension), and via
Monte Carlo simulation using system sizes of up to 5000 sites.  The
latter yield the estimates $p_c = 0.92965(3)$, $\beta/\nu_\perp =
0.247(2)$, $z = \nu_{||}/\nu_\perp = 1.45(3)$ and $\beta =
0.412(4)$. A similar value, $\beta = 0.42(1)$, was obtained in Ref.
\cite{mancam} using a series of cluster approximations (of up to 11
sites), combined with Suzuki's coherent anomaly analysis
\cite{suzukicam}.

The studies reported here employ the quasistationary (QS) simulation
method, which, due to increased efficiency in the critical region,
permits a tenfold increase in the system size as compared to Ref.
\cite{mnrst}.  The QS method, described in detail in \cite{qsspre},
provides a just sampling of asymptotic (long-time) properties,
conditioned on survival.  In practice this is accomplished by
maintaining (and gradually updating) a set of configurations visited
during the evolution; when a transition to the absorbing state is
imminent the system is instead placed in one of the saved
configurations.  Otherwise the evolution is exactly that of a
``standard" simulation algorithm such as used in Ref. \cite{mnrst}.

\section{Simulation results}

I performed two sets of studies using the QS method.  The first is
used to determine the QS order parameter (defined as the faction
$\rho$ of active sites), the moment ratio $m = \langle \rho^2
\rangle/\rho^2$, and the mean lifetime $\tau$ of the quasistationary
state, in the immediate vicinity of the critical point $p_c$, for
system sizes $L= 1000$, 2000, 5000, 10$\,$000, 20$\,$000 and
50$\,$000.  (The QS lifetime is taken as the mean number of time
steps between successive attempts to visit the absorbing state.) A
second set of simulations is used to study the supercritical regime
($p > p_c$) for system sizes $L=10 \, 000$, 20$\,$000 and 50$\,$000.
(For $p$ substantially larger than $p_c$, the lifetime is much
larger than the simulation time, so that the system never visits the
absorbing state, and the QS method becomes identical to a standard
simulation.)

Each realization of the process is run for 10$^9$ time steps;
averages are taken in the QS regime, which necessitates discarding
an initial transient that ranges from 10$^6$ time steps (for
$L=1000$) to 10$^8$ time steps (for $L=50 \, 000$). The number of
saved configurations ranges from 1000 (for $L=1000$) to 400 (for
$L=50\, 000$).  The list updating probability $p_{rep}$ ranges from
$10^{-3}$ (for $L=1000$) to $5 \times 10^{-6}$ (for $L=50 \, 000$).
During the initial relaxation period $p_{rep}$ is increased by a
factor of ten to erase the memory of the initial configuration.

I first discuss the studies focusing on the critical region. As in
Ref. \cite{mnrst}, I study, for each system size, a series of
particle number values $N$, chosen so that $p = N/L$ lies
immediately above or below $p_c$. Since the particle density can
only be varied in steps of $1/L$, estimates for properties at
intermediate values of $p$ are obtained via interpolation. The
results of the QS simulations were found to agree, to within
uncertainty, with the corresponding results of conventional
simulations \cite{mnrst}, for $L = 1000$, 2000 and 5000.    The
criterion for criticality is power-law dependence of $\rho$ and
$\tau$ on system size, i.e., the familiar relations $\rho \sim
L^{-\beta/\nu_\perp}$ and $\tau \sim L^z$, and constancy of the
moment ratio $m$ with $L$. The most sensitive indicator turns out to
be the order parameter $\rho$. Using the data for system sizes 5000
- 50$\,$000, I rule out $p$ values that yield a statistically
significant curvature of the graph of $\ln \rho$ versus $\ln L$.
This results in the estimate $p_c = 0.929780(7)$. (For the remainder
of the analysis $p_c$ is fixed at this value and is no longer
available as an adjustable parameter.) The associated exponent is
$\beta/\nu_\perp = 0.213(6)$, where the uncertainty represents a
contribution ($\pm 0.005$) due to the uncertainty in $p_c$ and a
small additional uncertainty in the linear fit to the data.
Simulation results for $\rho$ as a function of $L$, for various
densities near $p_c$, are shown in Fig. 1; curvature of the plots
for off-critical values is evident in the inset.

The data for the QS lifetime $\tau$ furnish a similar but somewhat
less precise estimate, $p_c = 0.929777(17)$.  Fitting the data for
$L=5000$ - 50$\,$000, using the $p_c$ interval obtained from the
analysis of $\rho$, I find $z = 1.50(4)$.  The moment ratio $m$ is
also useful for setting limits on $p_c$.  As shown in Fig. 2, this
quantity appears to grow with system size for $p< p_c$ and
vice-versa; we may rule out the values 0.92976 and 0.92980 on this
basis. The moment ratio data yield $m_c = 1.142(8)$. The main
contribution to the uncertainties in $z$ and $m$ is again due to the
uncertainty in $p_c$.

The present estimate for $p_c$ is significantly greater than that
found in Ref. \cite{mnrst}, although the difference amounts to about
0.01\%.  The results for the exponent $z$ are consistent, but the
present study yields a substantially (16\%) lower estimate for
$\beta/\nu_\perp$ than reported previously. The present result for
$m_c$ is also substantially lower than the value 1.1596(4) reported
in Ref. \cite{mnrst}.  These differences highlight the strong
finite-size corrections affecting stochastic sandpiles.

I turn now to the results for the order parameter in the
supercritical regime.  Fig. 3 shows that the data for system sizes
10$\,$000, 20$\,$000 and 50$\,$000 are well converged for $\Delta =
p - p_c \geq 10^{-3}$, that is, finite-size effects are only present
nearer the critical point.  Evidently, the data are not consistent
with a simple power law of the form $\rho \sim \Delta^\beta$. Indeed
this departure from the familiar behavior of the order parameter was
already noted (with data for smaller systems) in Ref. \cite{mnrst}.
In the latter work the power law was ``restored" by introducing a
size-dependent critical density $p_c(L) \simeq p_{c,\infty} -
Const/L^{1/\nu_\perp}$, leading to a series of estimates for the
critical exponent $\beta$ that increase systematically with $L$,
apparently converging to $\beta = 0.412(4)$.  With the present data,
which are converged over a broader range of $\Delta$ values, I find
that shifting the critical value {\it does not lead to an apparent
power law}.

One is therefore left to conclude that either the order parameter
does not obey power-law scaling, or that there are unusually strong
corrections to scaling.  Including a correction to scaling term, one
has

\begin{equation}
\rho \sim \Delta^\beta \left[ 1 + A \Delta^{\beta'} \right]
\label{cts}
\end{equation}
so that there are now three adjustable parameters, $\beta$, $\beta'$
and $A$.  Even with a reasonably large number of data points (18 for
$L = 10\, 000$), this induces a huge range of variation in the
exponent $\beta$.  Decent fits can be obtained with values as low as
$\beta = 0.1$ and as large as 0.3.

To resolve this difficulty I return to the data in the immediate
vicinity of $p_c$.  These data can be used to determine the
correlation length exponent $\nu_\perp$ in the following manner.
Finite-size scaling implies that for $p \simeq p_c$, the moment
ratio obeys the relation

\begin{equation}
m(\Delta, L) \simeq {\cal F}_m (L^{1/\nu_\perp} \Delta).
\label{fssm}
\end{equation}
where ${\cal F}_m$ is a scaling function.  This implies that

\begin{equation}
\left| \frac{\partial m}{\partial p} \right|_{p_c} \propto
L^{1/\nu_\perp}.
\label{fssderiv}
\end{equation}
Moreover, the finite-size expression $\rho = L^{-\beta/\nu_\perp}
{\cal F}_\rho (L^{1/\nu_\perp} \Delta)$ implies that

\begin{equation}
\left| \frac{\partial \ln \rho}{\partial p} \right|_{p_c} \propto
L^{1/\nu_\perp},
\label{fssderivrho}
\end{equation}
and similarly for the derivative of $\ln \tau$ at the critical
point.  The derivatives are evaluated numerically as follows. For
each value of $L$ studied, data for five values of $p$ clustered
around $p_c$ are fit with a cubic polynomial; the derivative of the
polynomial is then evaluated at $p_c$. The resulting derivatives are
plotted in Fig. 4; clean power laws are observed, leading to
$\nu_\perp = 1.362(7)$, 1.323(14) and 1.372(21), using the data for
$\ln \rho$, $m$ and $\ln \tau$, respectively.  Pooling these results
yields the estimate $\nu_\perp = 1.355(18)$.  Then, using the values
for $\beta/\nu_\perp$ and $z$ reported above, I find $\beta =
0.289(12)$ and $\nu_{||} = 2.03(8)$.

Using this value for $\beta$, the data for the order parameter in
the supercritical regime can be fit using the correction to scaling
form, Eq. \ref{cts}, with parameters $\beta' = 0.446$ and $A =
1.3505$.  For $\Delta = 0.1$, the correction term $A
\Delta^{\beta'}$ in Eq. \ref{cts} is $0.48$, showing that there are
sizeable deviations from a pure power law.  It is usual to verify
scaling by seeking a data collapse, plotting $\rho^* =
L^{\beta/\nu_\perp} \rho$ versus $\Delta^* = L^{1/\nu_\perp}
\Delta$.  For $\Delta > 0.001$ the order parameter does not follow a
pure power law and so the data cannot collapse.  It is nevertheless
of interest to construct such a scaling plot (Fig. 5). Although the data do
not collapse over most of the range, they do collapse in the
interval $-1 \leq \Delta \leq 1$.  A linear fit to the data in this
interval yields a slope of 0.27(1).  This is close to the $\beta$
value obtained from the finite-size scaling analysis, suggesting
that simple scaling is restricted to a narrow interval very near the
critical point.

\section {Discussion}

A study of the one-dimensional restricted-height stochastic sandpile
using quasistationary simulations permits study of systems an order
of magnitude larger than previously studied, and yields critical
properties different than those obtained previously.  In the case of
the critical density, the small change (about 0.01\%) from the
previous estimate may be attributed to finite-size effects, which
are known to affect sandpile models strongly.

Of greater concern are critical exponent values, since they define
the universality class of the model.  Since there is every reason
(based on symmetry considerations) to expect the restricted sandpile
to belong to the same universality class as the unrestricted version
(indeed, this seems well established in two dimensions
\cite{mnrst}), I collect, in Table I, critical exponent values from
various studies of stochastic sandpiles, C-DP and the conserved
threshold transfer process (CTTP), also expected to belong to the
same class.

The overall conclusion from Table I is that studies using smaller
lattices yield values in the range 0.38 - 0.42 for the exponent
$\beta$ (Ref. \cite{lubeckcttp} is however an exception), and that
the large-scale simulation of Ref. \cite{kockelkoren}, the numerical
study of the C-DP field theory \cite{ramasco} and the present work
yield a consistent set of results, with $\beta \simeq 0.29$. (A
similar value has been found for a modified conserved lattice gas
model \cite{deoliveira}.) Although the system size (4000 sites) used
in the field theory simulations is not large, one should note that
each `site' in such a simulation may represent a region comprising
many lattice sites in the original model.  Compared with the earlier
sandpile simulations, the distinctive feature of the present work
may not be system size, but the fact that here the exponent $\beta$
is determined via finite-size scaling {\it at the critical point},
rather than from the usual analysis of the order parameter in the
supercritical regime. Indeed, it is easy to see from Fig. 3 that
data for $\Delta = p - p_c$ in the range $10^{-3}$ - $10^{-1}$ will
yield larger estimates for $\beta$.  (The same observation applies
to the CAM analysis \cite{mancam}, which essentially probes the
shape of the function $\rho(\Delta)$ at some distance from the
critical point $\Delta = 0$.) I observe a simple power-law behavior,
and data collapse for various lattice sizes, only in a restricted
range of the scaling variable $\Delta^* = L^{1/\nu_\perp} \Delta$.

Also included in Table I are exponent values for one-dimensional
directed percolation \cite{jensendp}.  The values obtained in Refs.
\cite{ramasco} and \cite{kockelkoren}, as well as in the present
work, are not very different from those of DP.  A clear difference
from DP scaling was however demonstrated in Ref. \cite{dornic},
where the initial decay exponent for one-dimensional C-DP is found
to be $\theta = 0.125(2)$, as opposed to $0.1595(1)$ for DP.  The
rather substantial differences found here in $\beta/\nu_\perp$, and
in the moment ratio $m$ (1.142(8) for the restricted sandpile
compared with 1.1736(1) for DP \cite{qsspre}), lend further support
to the conclusion that the C-DP/stochastic sandpile universality
class is distinct from that of directed percolation, as is evidently
the case in two dimensions. (This despite the result \cite{mohanty},
that when suitably modified to include `sticky grains', sandpiles
fall generically in the DP class.)

In summary, I have applied the quasistationary simulation method to
a one-dimensional restricted-height stochastic sandpile, and obtained
results consistent with recent studies of C-DP.  This supports the
assertion that the latter class includes stochastic sandpiles, as
would be expected on the basis of symmetry and conservation laws.

\vspace{2em}

\noindent{\bf Acknowledgements}
\vspace{1em}

I am grateful to Hugues Chat\'e, M\'ario de Oliveira and Miguel
Angel Mu\~noz for helpful discussions and comments on the
manuscript. This work was supported by CNPq and Fapemig, Brazil.

\newpage

\begin{table}[h]
\caption{\sf Summary of exponent values for one-dimensional models
in the C-DP universality class.  $L_{max}$ denotes the largest
system size studied. Abbreviations: CAM: coherent anomaly method;
FT: field theory.}
\begin{center}
\begin{tabular}{|l|r|c|c|c|}
\hline
Model& $L_{max}$ & $\beta$ &  $\beta/\nu_\perp$ & $z$ \\
\hline\hline
Manna \cite{manna1d}      & 10$\,$000  & 0.42(2) & 0.24(1)    & 1.66(7)   \\
Manna \cite{lubeckheger}  &  8192      &         & 0.28(3)    & 1.39(11) \\
CTTP  \cite{lubeckcttp}   & 131$\,$072 & 0.38(2) & 0.24(1)    & 1.66(7)   \\
Rest. Manna \cite{mnrst}  & 5000       & 0.416(4)& 0.246(5)   & 1.50(9)  \\
Rest. Manna CAM \cite{mancam} &      & 0.41(1) &            &            \\
C-DP \cite{kockelkoren}  & 4.2 $\times 10^6$ & 0.29(2) &            & 1.55(3)   \\
C-DP FT  \cite{ramasco}   & 4000       & 0.28(2) & 0.214(8)   & 1.47(4)   \\
Rest. Manna (present work) & 50$\,$000 & 0.289(12) & 0.213(6) &
1.50(4)
\\ \hline
DP  \cite{jensendp}       &            & 0.2765  & 0.2521     & 1.5807      \\
\hline
\end{tabular}
\end{center}
\end{table}

\newpage
\noindent FIGURE CAPTIONS
\vspace{1em}

\noindent FIG. 1.  Stationary order parameter versus system size for
particle densities (bottom to top) $p=0.92977$, 0.92978 and 0.92979.
Inset: $\ln L^{0.213} \rho$ versus $\ln L$ for the same set of
particle densities. \vspace{1em}

\noindent FIG. 2.  Moment ratio $m$ versus system size for particle
densities (top to bottom) $p=0.92976$, 0.92978 and 0.92980.
\vspace{1em}

\noindent FIG. 3.  Stationary order parameter versus $\Delta = p -
p_c$ for system sizes (top to bottom) $L=10^4$, $2 \times 10^4$ and
$5 \times 10^4$. \vspace{1em}

\noindent FIG. 4.  Derivatives of (lower to upper) $\ln \tau$, $\ln
\rho$ and $m$ with respect to particle density, evaluated at $p_c$,
versus system size.  The slope of the straight line is 0.734.
\vspace{.5em}

\noindent FIG. 5.  Scaled density $\rho^*$ versus scaled distance
from critical point $\Delta^*$, as defined in text.  System sizes:
$10^4$ (open squares); $2 \times 10^4$ (filled squares); $5 \times
10^4$ (diamonds).

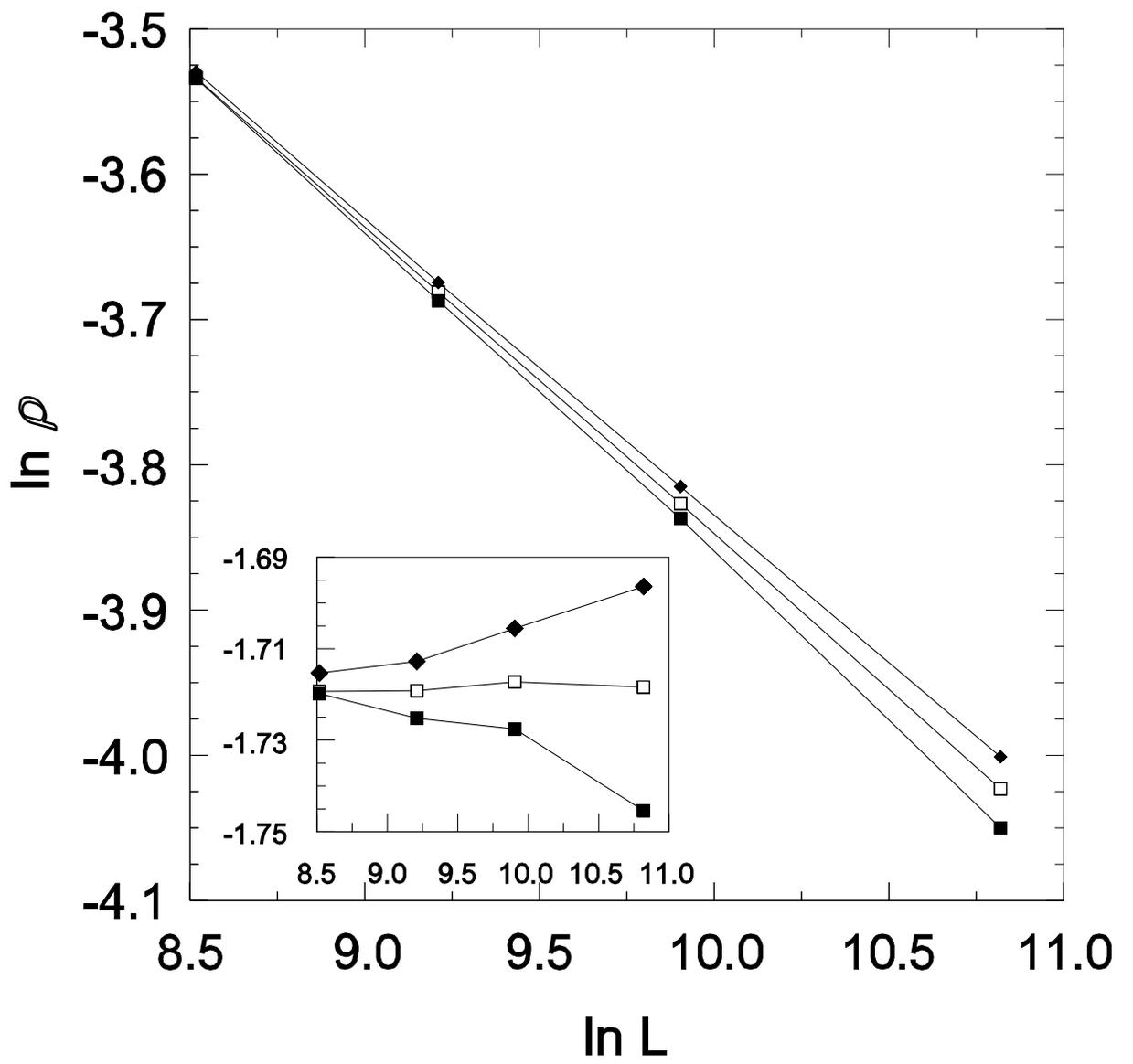
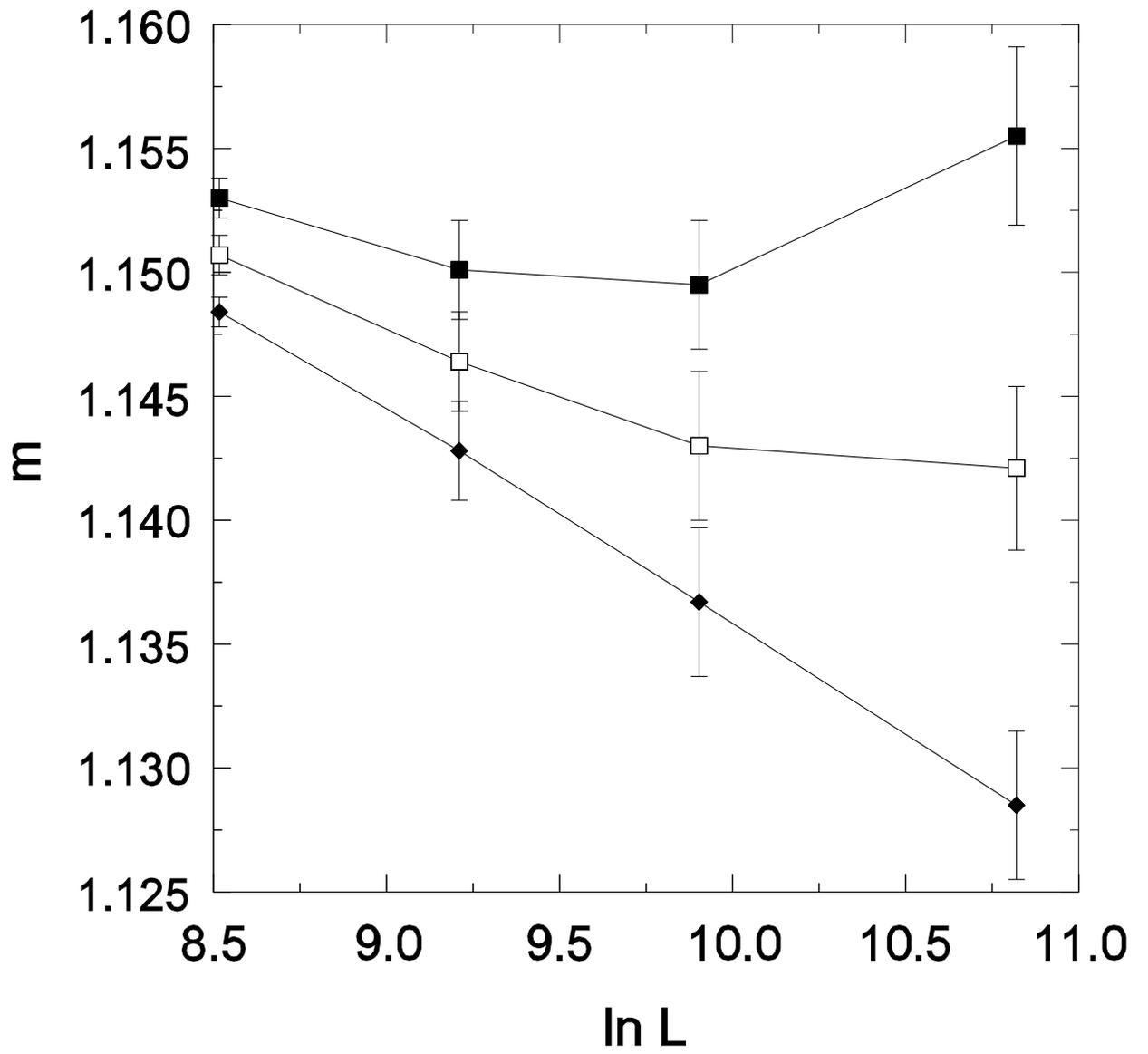
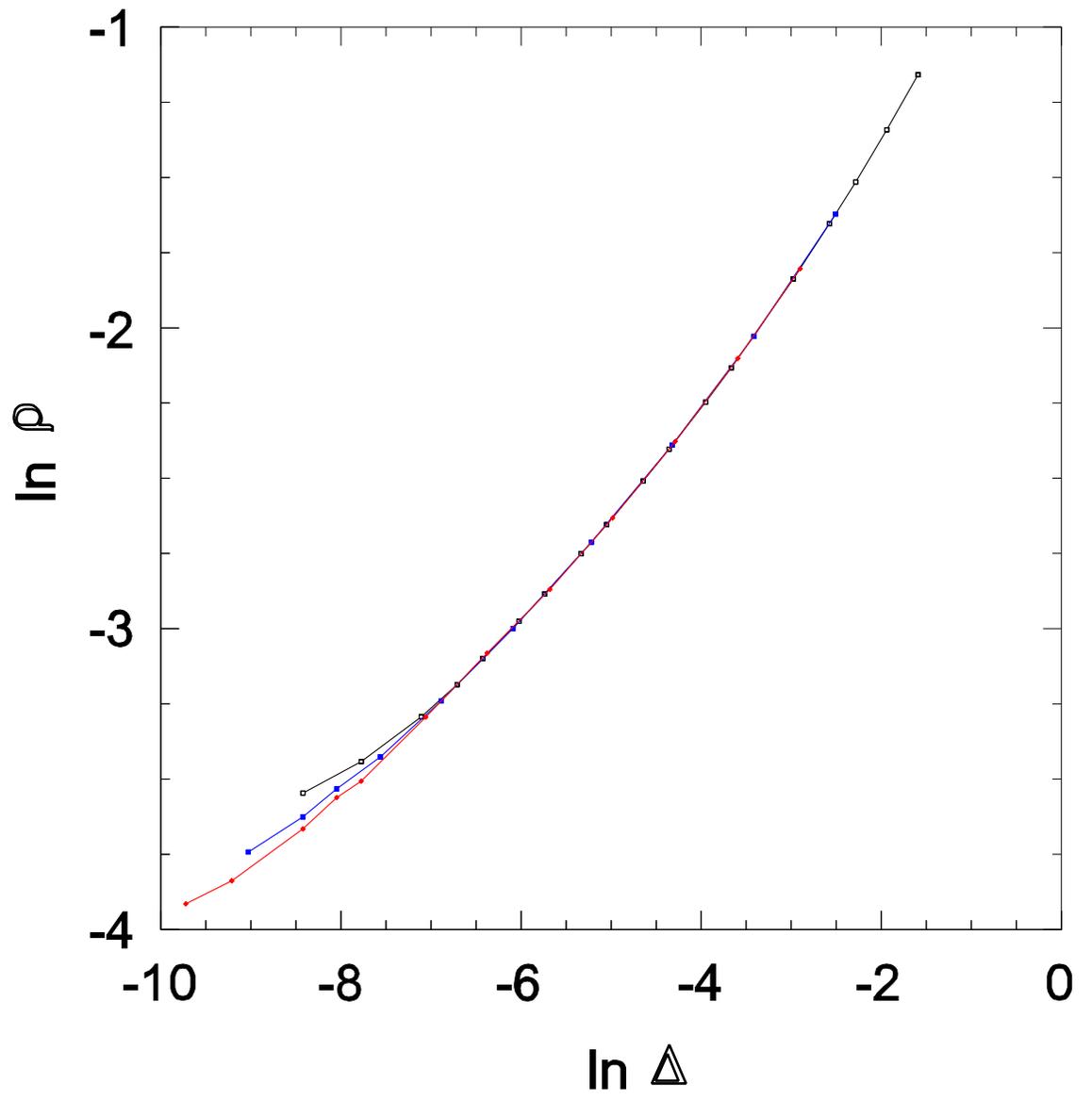
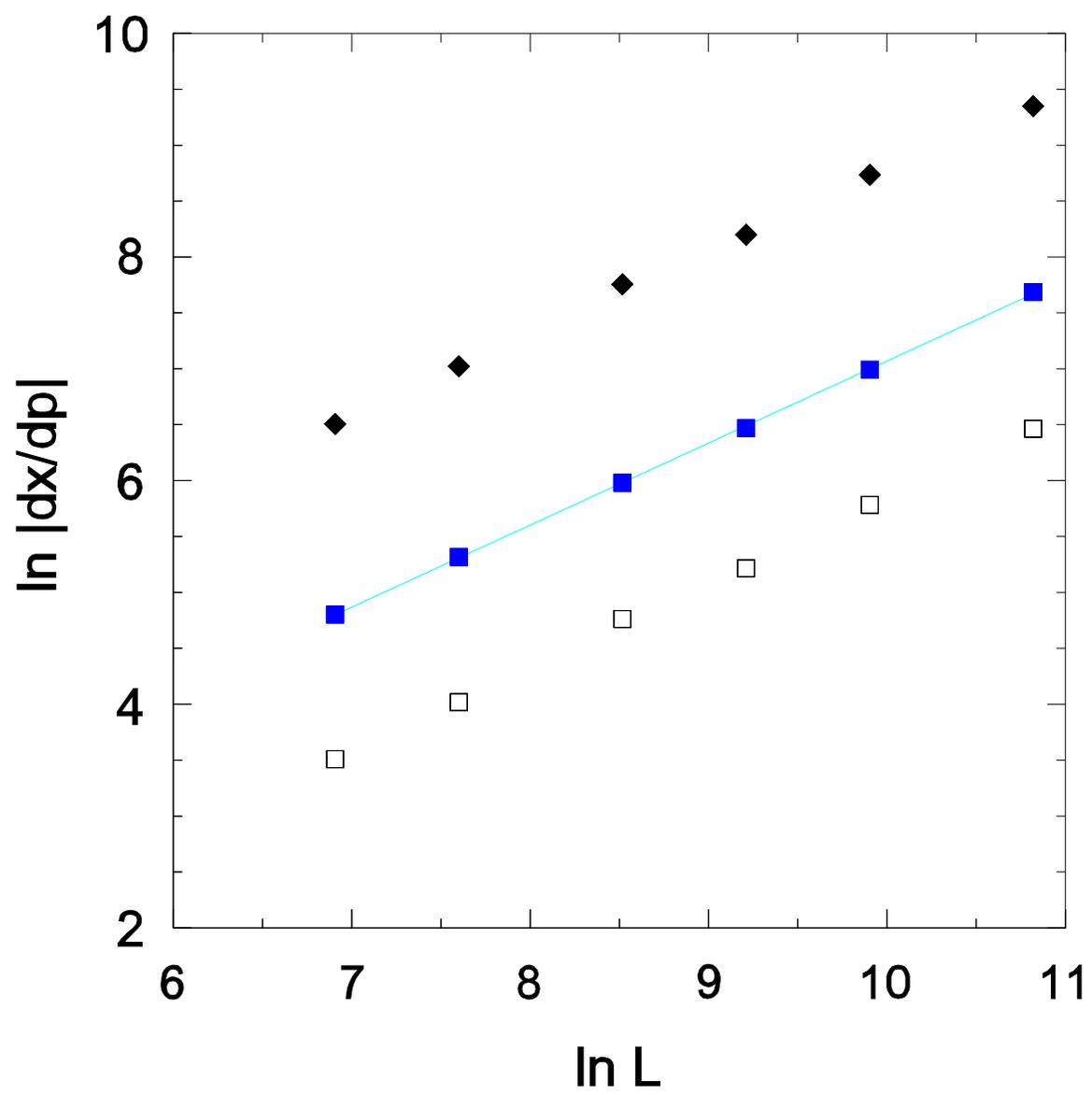
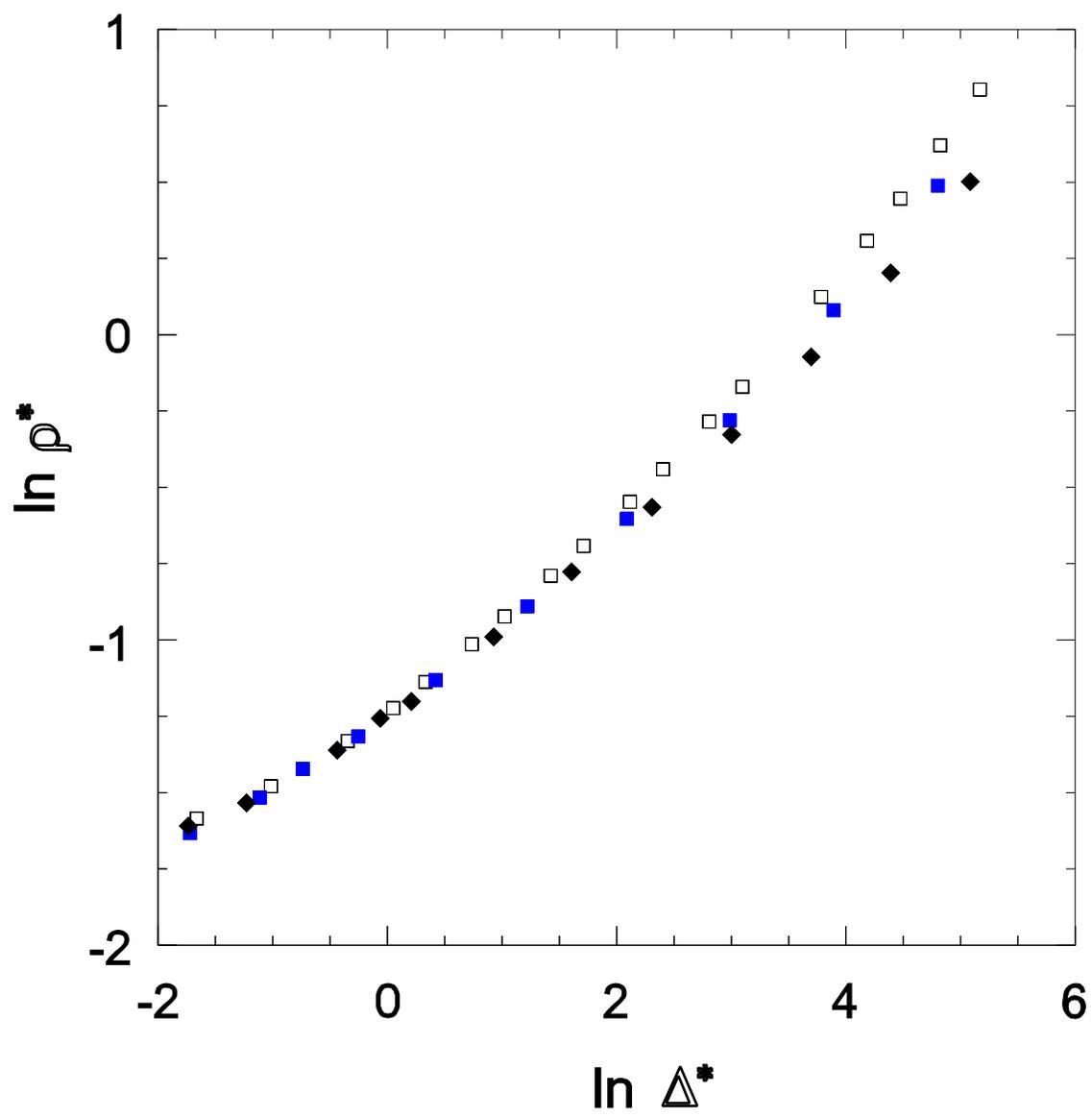
\end{document}